\documentclass{PoS}

\usepackage{epsfig}
\usepackage{amsmath}
\usepackage{dsfont}

\newcommand{\beq}{\begin{equation}}
\newcommand{\eeq}{\end{equation}}
\newcommand{\bea}{\begin{eqnarray}}
\newcommand{\eea}{\end{eqnarray}}

\newcommand{\tr}{\hbox{tr}}

\newcommand\eqn[1]{(\ref{#1})}      
\newcommand\Eqn[1]{Eq.~(\ref{#1})}  
\newcommand\Fig[1]{Fig.~\ref{#1}}  

\title{A perturbative description of the deconfinement transition in Yang-Mills theories}

\ShortTitle{Perturbative deconfinement transition}

\author{\speaker{Julien Serreau}\thanks{Based on work in collaboration with U. Reinosa, M. Tissier, and N. Wschebor.}\\
        Astro-Particule et Cosmologie (APC), CNRS UMR 7164, Universit\'e Paris Diderot\\ 10, rue Alice Domon et L\'eonie Duquet, 75205 Paris Cedex 13, France.\\
        E-mail: \email{serreau@apc.univ-paris7.fr}}

\abstract{We investigate the deconfinement transition of static quarks in SU($N$) Yang-Mills theories using a perturbative approach based on a massive extension of the Landau-DeWitt gauge-fixed action, where the gluon mass term is related to the issue of Gribov ambiguities. A leading-order, one-loop calculation of the effective potential for the Polyakov loop produces a deconfinement transition of second order for the SU($2$) theory and of first order for SU($3$) with transition temperatures in qualitative agreement with known values. We also report on the results of a two-loop calculation of the critical temperature and of thermodynamical quantities in the SU($2$) case.}

\FullConference{9th International Workshop on Critical Point and Onset of Deconfinement - CPOD2014,\\
		17-21 November 2014\\
		ZiF (Center of Interdisciplinary Research), University of Bielefeld, Germany}

\begin{document}

\section{Introduction: Massive (gauge-fixed) gluons and Gribov ambiguities}

Progress in the understanding of the phase structure of QCD is mainly driven by lattice results. It is, by now, well established that there is a crossover from a hadron phase to a quark-gluon plasma phase along the temperature axis \cite{Petreczky:2012rq,Borsanyi-thisproc}. In contrast, a simple perturbative description of the confinement-deconfinement transition seems out of reach even in the case of pure Yang-Mills theories because weak coupling techniques break down at temperatures of the order of the transition temperature. Continuum calculations are either based on nonperturbative approaches such as the functional renormalization group (FRG) or Dyson-Schwinger equations (DSE) \cite{Fischer:2009wc,Fister:2013bh,Fisher-thisproc}, or on the use of phenomenological models \cite{Pisarski:2000eq,Dumitru:2012fw}. 

However, the failure of standard perturbative techniques at low temperatures should be interpreted with care due to the Gribov problem. Indeed, the Faddeev-Popov (FP) quantization procedure ignores the Gribov ambiguities inherent to the issue of fixing a gauge in a continuous way in a nonabelian theory \cite{Gribov77} and is, at best, a good approximation at high energies, where the Gribov ambiguities play a negligible role. A fully consistent quantization procedure must take into account the Gribov problem. A well-established example is the so-called minimal Landau gauge in the context of lattice gauge-fixed calculations, which consists in selecting one (arbitrary) Gribov copy on each gauge orbit. Explicit calculations of the Euclidean vacuum propagators in this gauge have revealed that the gluon behaves as a massive field in the regime of deep infrared momenta, whereas the ghost remains massless \cite{Boucaud:2011ug}. This demonstrates that the BRST symmetry of the FP action---which prohibits a gluon mass or, more precisely, a nonzero value of the inverse gluon propagator at zero momentum---is not realized in a fully gauge-fixed setting. In fact, it is also well-known that lattice implementations of the BRST symmetry typically lead to ill-defined zero over zero expressions for physical observables \cite{Neuberger:1986vv}. 

Unfortunately, the minimal Landau gauge cannot be implemented in terms of a local renormalizable action and is not suited for continuum approaches. An early attempt to go beyond the FP quantization consists in restricting the path integral over gauge fields to the first Gribov region, where the  FP operator is positive definite \cite{Gribov77,Zwanziger89}. The original proposal predicts a vanishing gluon propagator at zero momentum, which agrees with lattice results in the Coulomb gauge \cite{Burgio:2008jr} but not with those in the Landau gauge mentioned above. A refined version, which includes various condensates of mass-dimension two gives results in agreement with the lattice data in the Landau gauge \cite{Dudal08}. Finite temperature effects have been investigated in this context in Refs. \cite{Zwanziger:2004np,Fukushima:2013xsa,Canfora:2013kma}.

Yet another approach to the Gribov problem, advocated in Refs.~\cite{Tissier_10,Weber:2012vf}, is to acknowledge the fact that the BRST symmetry is likely not to be realized in a consistent gauge fixing procedure and to consider the minimal extension of the FP action consistent with locality and renormalizability. In the case of the Landau gauge, this is given by the Landau limit of the Curci-Ferrari (CF) model \cite{Curci76} which corresponds to the FP action augmented by a bare gluon mass term. One-loop calculations of the vacuum ghost and gluon propagators in this model have been shown to be in quantitative agreement with the aforementioned lattice data \cite{Tissier_10}. Moreover, Tissier and Wschebor have shown that the model admits infrared-safe renormalization group trajectories, where the coupling constant remains finite at all scales, which allows one to push perturbative calculations down to deep infrared momenta. The gluon mass regulates the infrared divergences responsible for the Landau pole in the FP theory. This approach has also been successfully applied to the one-loop calculation of three-point vacuum correlators in Yang-Mills theories \cite{Pelaez:2013cpa}, to the vacuum propagators of QCD \cite{Pelaez:2014mxa} and to the ghost and gluon propagators of Yang-Mills theories at finite temperature \cite{Reinosa:2013twa}.

Finally, the relation of this phenomenological approach to the issue of Gribov ambiguities has been given a more solid theoretical foundation in Ref.~\cite{Serreau:2012cg}, where a novel quantization procedure for Yang-Mills theories was put forward, which consists in taking a particular average over all Gribov copies instead of trying to select a unique one. This avoids the Neuberger 0/0 problem mentioned above. For a suitably chosen averaging procedure, this can be formulated as a local action which is perturbatively  renormalizable in four spacetime dimensions. In the case of the Landau gauge\footnote{A more general class of (nonlinear) covariant gauges has been considered in Ref.~\cite{Serreau:2013ila}.} and for a suitable definition of the renormalized mass parameter, the resulting gauge-fixed theory has been shown to be perturbatively equivalent to the CF model for the calculation of ghost and gluon correlation functions. The bare gluon mass is related to the averaging weight which lifts the degeneracy between Gribov copies. 

The results mentioned above in the CF model show that this approach leads to a better-behaved perturbation theory than the usual FP procedure. A qualitative but instructive analogy can be made with the physics of the Kosterlitz-Thouless transition in the $XY$ model in statistical physics \cite{Kosterlitz:1973xp}. In terms of usual spin waves, the physics of the transition is intrinsically nonperturbative: perturbation theory shows no sign of the transition at all orders. Alternatively, taking explicitly into account the vortex excitations of the $XY$ model leads to a valid perturbative description of the transition. 

These considerations have motivated the works of Refs.~\cite{Reinosa:2014ooa,Reinosa:2014zta}, summarized below, where the confinement-deconfinement transition of pure SU($N$) Yang-Mills theories at finite temperature is studied in the context of the modified (massive) perturbative approach described above. A crucial ingredient is to devise a loop expansion which correctly captures the $Z_N$ center symmetry of the theory, which is spontaneously broken in the deconfined phase. This is easily achieved in the context of background field methods. In particular, we consider the minimal massive extension of the standard Landau-DeWitt gauge, the background field generalization of the Landau gauge. A one-loop calculation of the effective potential for the Polyakov loop---the order parameter of the $Z_N$ transition---correctly describes a confined phase at low temperatures and a deconfinement transition of second and of first order for the cases $N=2$ and $N=3$ respectively, with transition temperatures in qualitative agreement with known values \cite{Reinosa:2014ooa}. Finally, we briefly review the work of Ref.~\cite{Reinosa:2014zta} concerning the calculation of the critical temperature as well as of the thermodynamic pressure and entropy of the SU($2$) theory at two-loop order.

\section{Massive Landau-DeWitt gauge}

We consider the Euclidean Yang-Mills action in $d=4$ dimensions\footnote{For simplicity, we ignore regularization and renormalization issues in this contribution. For details, see Ref.~\cite{Reinosa:2014zta}.} 
\beq 
\label{eq:YM}
S_{\rm  YM}=\frac{1}{2}\int_x \tr\left\{F_{\mu\nu}F_{\mu\nu}\right\}\,, 
\eeq 
where $F_{\mu\nu}=\partial_\mu A_\nu-\partial_\nu A_\mu -ig[A_\mu,A_\nu]$, with $g$ the coupling constant, and $A_\mu=A_\mu^at^a$, with SU($N$) generators $t^a$ normalized as ${\rm tr}(t^at^b)=\delta^{ab}/2$. Finally, $\int_x\equiv\int_0^\beta d\tau\int d^{3} x$, with $\beta=1/T$ the inverse temperature. We quantize the theory using the background field method \cite{Weinberg:1996kr} with a background field $\bar A_\mu=\delta_{\mu0}\bar A_0^kt^k$ in the temporal direction and in the Cartan subalgebra spanned by the generators $t^k$, with $k=1,\ldots,N-1$. Our gauge-fixed action reads 
\beq
\label{eq_gf}
 S=\int_x\tr\left\{{1\over2}F_{\mu\nu}F_{\mu\nu}+{m^2}a_\mu a_\mu+2\bar D_\mu\bar cD_\mu c+2ih\bar D_\mu a_\mu\right\},
\eeq
with $a_\mu=A_\mu-\bar A_\mu$ the fluctuating gluon field, $h$ a (real) Nakanishi-Lautrup field and $c$ and $\bar c$ the FP ghost and antighost fields. Here,  $\bar D_\mu\varphi=\partial_\mu\varphi -ig [\bar A_\mu,\varphi]$ and  $D_\mu\varphi=\partial_\mu\varphi -ig [A_\mu,\varphi]$. Apart from the bare gluon mass term $\propto m^2$, \Eqn{eq_gf} is nothing but the FP action corresponding to the Landau-DeWitt gauge: $ \bar D_\mu a_\mu=0$. The relation of the gluon mass term in \Eqn{eq_gf} with the Gribov ambiguities of this gauge condition is discussed in Ref.~\cite{Reinosa:2014ooa}.

To evaluate physical observables at zero sources, one can minimize the following background field effective potential
\beq
\label{eq:poteff}
 V(T,r^k)=\frac{\Gamma[\bar A,\varphi=0]}{\beta \Omega}-V_{\rm vac},
\eeq
 where $r^k= \beta g\bar A_0^k$, $\Omega$ is the spatial volume, and where we have subtracted the zero temperature contribution $V_{\rm vac}$. Here, $\Gamma[\bar A,\varphi]$ is the effective action in presence of the background, with $\varphi=(a_\mu,c,\bar c,ih)$. 
The loop expansion of the potential \eqn{eq:poteff} corresponds to an expansion in powers of $g$ with $g\bar A_0\sim{\cal O}(1)$. We write the corresponding series as
\beq
\label{eq:pertexp}
 V(T,r^k)=\sum_{n\ge0}V^{(n)}(T,r^k)\,,
\eeq
with $\smash{V^{(n)}\sim{\cal O}(g^{2n-2})}$ the $n$-loop order contribution. The tree-level contribution vanishes identically, $V^{(0)}(T,r^k)=0$, and the first nontrivial term is the one-loop contribution.

\section{One-loop results}

Remarkably, the leading-order, one-loop contribution $V^{(1)}(T,r^k)$ already captures the essential physics of the $Z_N$ transition. This arises from the interplay between the respective contributions of the massive (gluon) and the massless (ghost) degrees of freedom. The one-loop expression reads
\beq
\label{eq:potoneloop}
 V^{(1)}(T,r^k)=\frac{1}{\beta \Omega}\left\{\frac{1}{2}{\rm Tr}\,\ln\,\Delta^{-1}_{ah}-{\rm Tr}\,\ln\,\Delta^{-1}_{c\bar c}\right\},
\eeq
with $\Delta^{-1}_{ah}$ and $\Delta^{-1}_{c\bar c}$ the tree-level inverse propagators in the $(a,h)$ and in the $(c,\bar c)$ sectors in the presence of the background field. The notation ${\rm Tr}$ on the right-hand side involves a trace over color and Lorentz indices as well as a sum over Matsubara frequencies and an integral over spatial momenta. The Matsubara sums are easily evaluated using standard contour integration techniques. Below we summarize the main results for the gauge groups SU($2$) and SU($3$) \cite{Reinosa:2014ooa}. 

\subsection{Second order phase transition for SU($2$)}

\begin{figure}[t!]  
\begin{center}
\epsfig{file=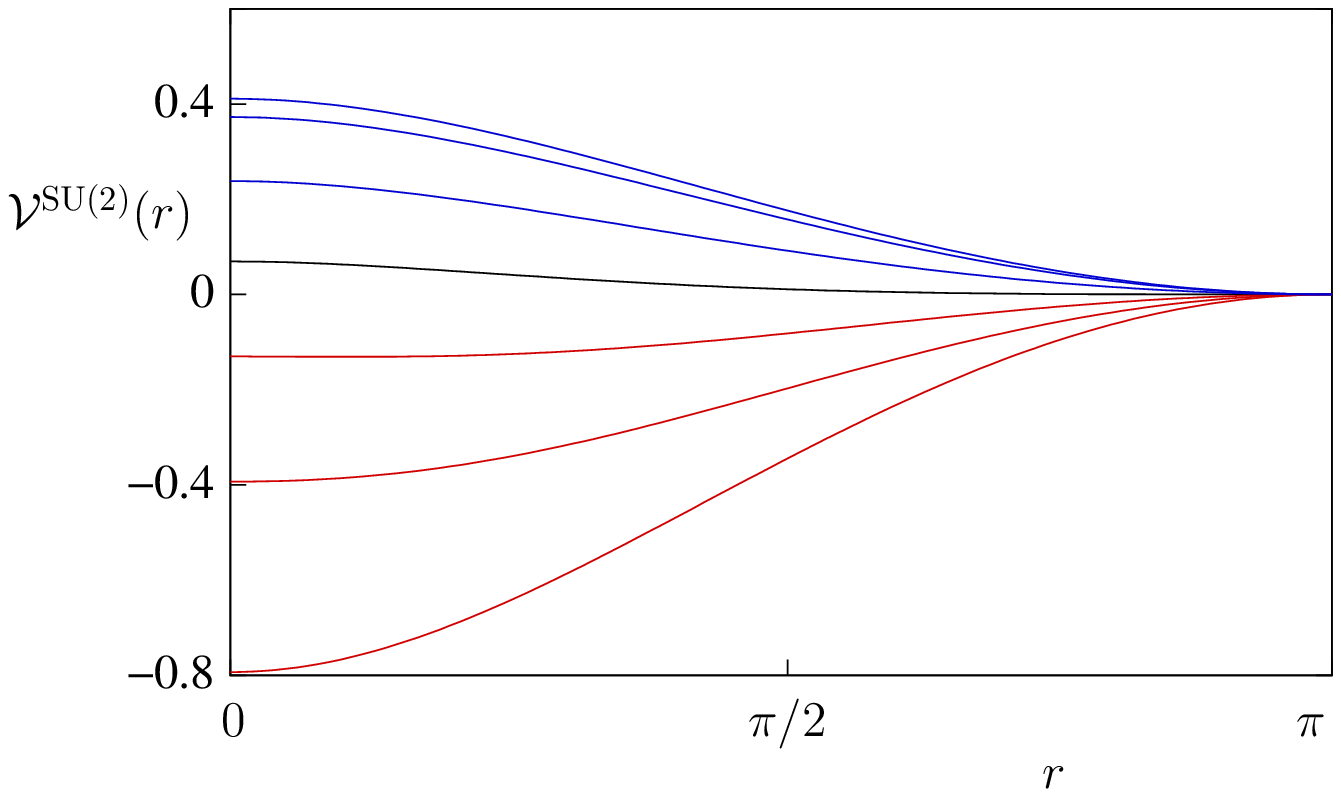,width=6.2cm}\qquad\quad
\epsfig{file=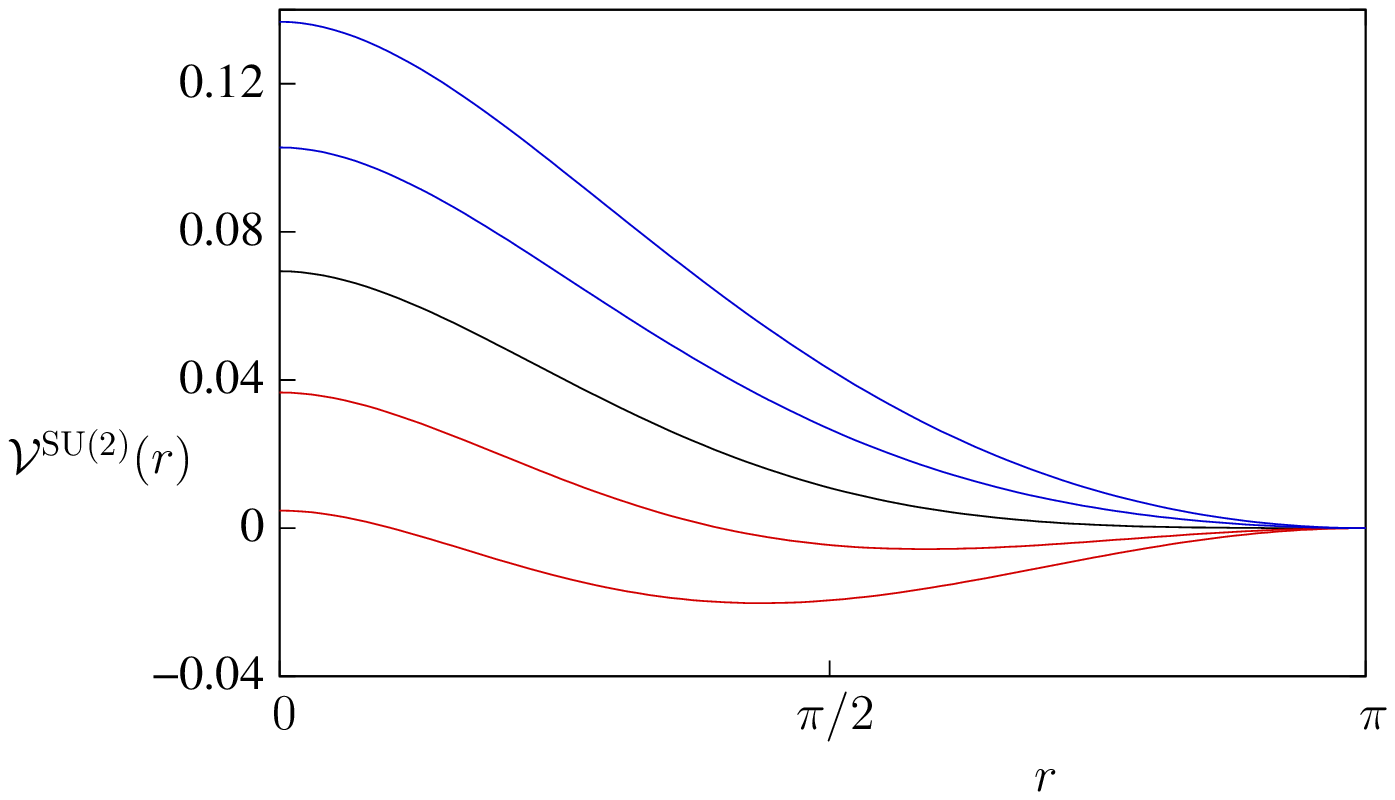,width=6.2cm}
 \caption{\label{fig:SU2pot} The SU($2$) dimensionless background field potential ${\cal V}(T,r)=V(T,r)/T^4$, normalized to its value at the confining point $r=\pi$, for temperatures $T=T_{c}$ (black), $T<T_{c}$ (blue), and $T>T_{c}$ (red). The minimum is at $r=\pi$ for $T\le T_c$ (black line) and continuously moves toward zero for higher temperatures. The left figure shows the Weiss and inverted Weiss potential at high and low temperatures respectively. The right figure is a close-up view around $T_c$.}
\end{center}
\end{figure}

In that case, the Cartan subalgebra has a single direction. Thanks to the symmetries of the problem one can reduce the analysis to the domain $r\in[0,\pi]$. Introducing the function
\beq\label{eq:Weiss0}
 {\cal F}_m(T,r)=\frac{T}{\pi^2}\int_0^\infty \!\!dq\,q^2\Big\{\ln\left(1-e^{-\beta\varepsilon_q}\right)+\ln\left(1+e^{-2\beta\varepsilon_q}-2e^{-\beta\varepsilon_q}\cos r\right)\Big\},
\eeq
which is such that, for $r\in[0,2\pi]$,
\beq\label{eq:Weiss}
{\cal F}_0(T,r)=\frac{T^4}{6}\left[\frac{(r-\pi)^4}{2\pi^2}-(r-\pi)^2+\frac{\pi^2}{10}\right],\nonumber\\
\eeq
the one-loop background field potential can be written as
\beq
\label{eq:onelooppot}
 V^{(1)}(T,r)=\frac{3}{2}{\cal F}_m(T,r)-\frac{1}{2}{\cal F}_0(T,r).
\eeq
The first term on the right-hand side is the contribution from the massive gluons and the second one is due to the incomplete cancelation of the massless modes in the $(a,h)$ and in the $(c,\bar c)$ sectors. At the same order of perturbation theory, the order parameter of the $Z_2$ transition, i.e., the average of the traced Polyakov loop is given by its tree-level expression
\beq
\label{eq:plo}
 \ell(T)=\cos\left(\frac{r_{\rm min}(T)}{2}\right),
\eeq
where $r_{\rm min}(T)$ is the absolute minimum of the one-loop potential \eqn{eq:onelooppot} at fixed temperature $T$.
In the high temperature limit, the effective gluon mass is negligible and one recovers the well-known Weiss potential \cite{Weiss:1980rj}, which corresponds to the standard FP quantization,
\beq\label{eq:one}
 V^{(1)}_{T\gg m}(T,r)\approx{\cal F}_0(T,r).
\eeq
The  minimum sits at $r=0$, corresponding to a deconfined phase, where $\ell(T)\neq0$. In contrast, at low temperatures the contribution from massive modes is exponentially suppressed and one obtains an inverted Weiss potential from the massless modes, as first discussed in Ref.~\cite{Fister:2013bh}:
\beq\label{eq:two}
 V^{(1)}_{T\ll m}(T,r)\approx-\frac{1}{2}{\cal F}_0(T,r).
\eeq
The minimum is now at $r=\pi$, which corresponds to a confined phase, with $\ell(T)=0$. A detailed study of the expression \eqn{eq:onelooppot} reveals a second order phase transition from the confined to the deconfined phase at a critical temperature $T_c/m\approx 0.33$, as illustrated in \Fig{fig:SU2pot}.

To estimate $T_c$, we use the value of the effective gluon mass parameter obtained by fitting the lattice data for the vacuum gluon propagator in the Landau gauge against the tree-level expression in the present theory at vanishing background. This is motivated by the fact that, for $T\le T_c$, the physical background is given by $\bar A_0=\pi T/g$ and thus vanishes at $T=0$. In $d=4$, we obtain $m=710$~MeV, which gives $T_c\simeq 238$~MeV. The typical lattice value \cite{Lucini:2012gg} is $T_c^{\rm latt}=295$~MeV and the most recent value from FRG/DSE studies\footnote{The authors of Ref.~\cite{Fister:2013bh} mention that their FRG result is modified to $T_c^{\rm FRG}=300\,{\rm MeV}$ when some backreaction effects---neglected in their main study---are included.} \cite{Fister:2013bh} is $T_c^{\rm FRG}=230$~MeV. Although such a comparison is only qualitative because of the issue of properly setting the scale, it is remarkable that the present one-loop result falls in the right ballpark.

\subsection{First order phase transition for SU($3$)}

\begin{figure}[t!]  
\begin{center}
\epsfig{file=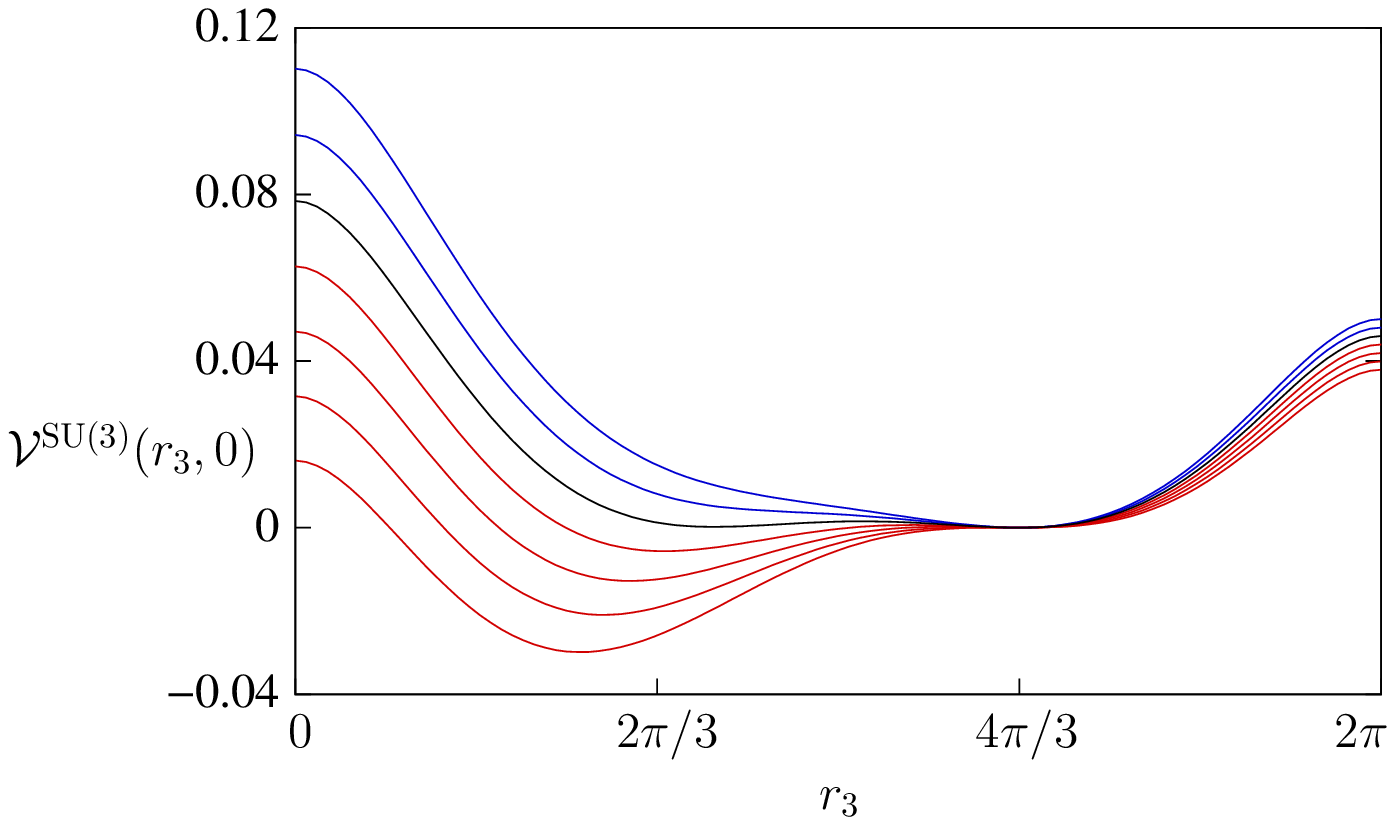,width=6.2cm}\qquad\quad
\epsfig{file=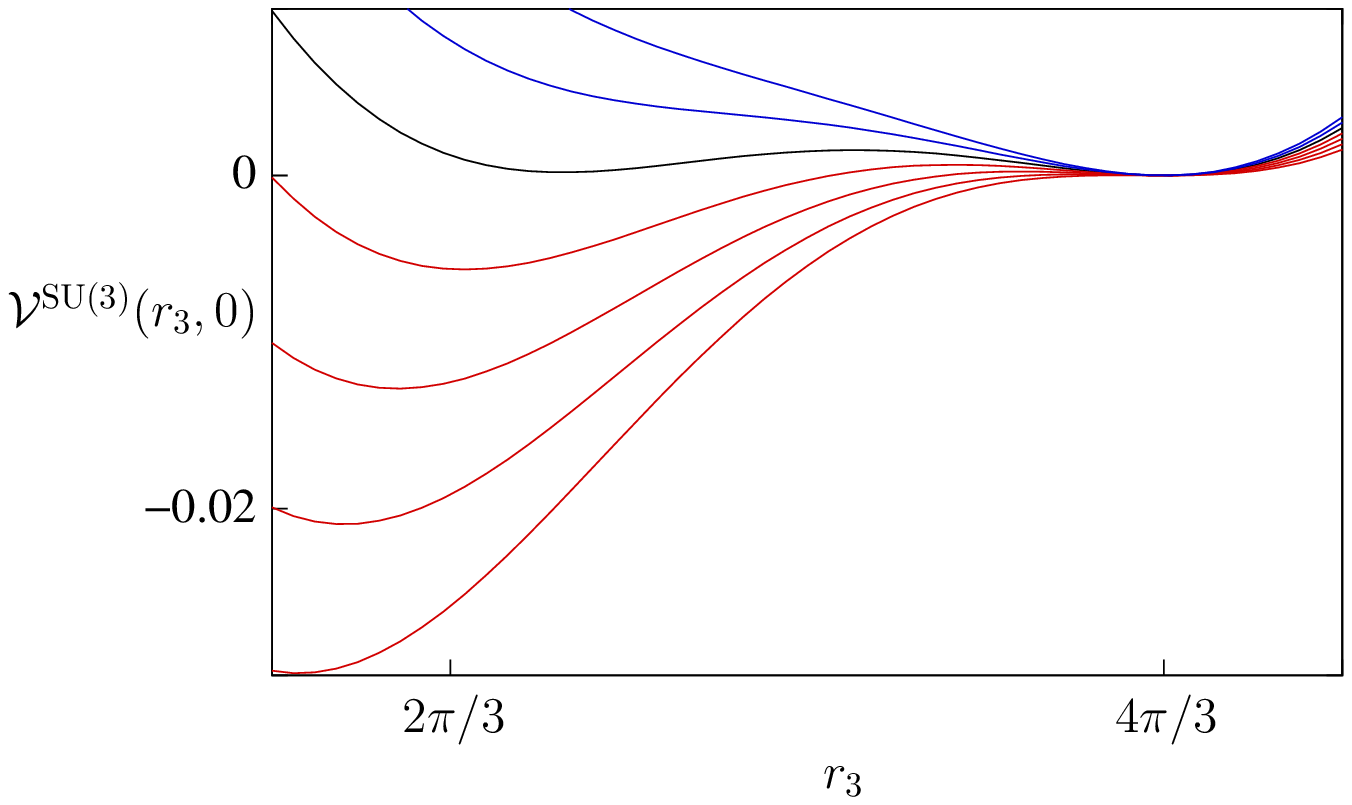,width=6.2cm}
\caption{\label{fig:SU3pot} The SU($3$) dimensionless background field potential ${\cal V}(T,r_3,r_8)=V(T,r_3,r_8)/T^4$ in the $r_8=0$ direction (where the absolute minimum always sits), normalized to its value at the confining point $(r_3=4\pi/3,r_8=0)$, for  temperatures $T=T_{c}$ (black), $T<T_{c}$ (blue), and $T>T_{c}$ (red). The right figure is a close-up view around $T_{\rm c}$. One observes a finite jump of the absolute minimum at $T=T_c$.}
\end{center}
\end{figure}

\begin{figure}[t!]  
\begin{center}
\epsfig{file=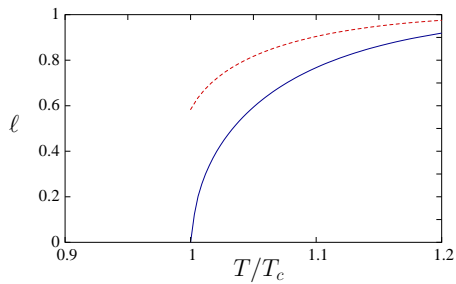,width=6cm}
 \caption{\label{fig:popol} The Polyakov loop as a function of the temperature normalized to the transition temperature for SU($2$) (blue) and SU($3$) (red), from the leading-order calculation.}
\end{center}
\end{figure}

The Cartan subalgebra has now two directions, conventionally taken as $(r_3,r_8)$. The calculation of the one-loop contribution \eqn{eq:potoneloop} is, again, elementary. As in the previous case, it reduces to the perturbative SU($3$) Weiss potential at high temperatures and to a confining, inverted Weiss potential at low temperatures. The transition is now of first order, as shown in \Fig{fig:SU3pot}, with a transition temperature $T_c/m\approx0.36$. As before, we estimate the gluon mass parameter from fits of SU($3$) lattice data for the vacuum propagator in the Landau gauge. We find $m=510$~MeV, which yields $T_c=185$~MeV. With the same word of caution as in the SU($2$) case, we mention the results from lattice calculations \cite{Lucini:2012gg}, $T_c^{\rm latt}=270$~MeV, and from FRG/DSE studies \cite{Fister:2013bh}, $T_c^{\rm FRG}=275$~MeV.
Our one-loop results are summarized in \Fig{fig:popol}, which shows the temperature dependence of the Polyakov loop for both SU($2$) and SU($3$). We mention that all the continuum approaches mentioned here produce a rapid increase of the Polyakov loop above $T_c$, in sharp contrast with the existing lattice data \cite{Dumitru:2012fw,Gupta:2007ax,Mykkanen:2012ri}.

\section{Two-loop results} 

An important aspect of the present perturbative approach is that the above leading-order results can be systematically improved by computing higher order corrections. The next-to-leading-order, two-loop contribution $V^{(2)}(T,r^k)$ to the background effective potential has been computed in Ref.~\cite{Reinosa:2014zta} for the SU($2$) theory. The calculation is considerably more involved than the leading-order one, but it can be done almost completely analytically, up to two-dimensional radial momentum integrals involving Bose-Einstein ditribution functions, which can be easily computed numerically. 

\begin{figure}[t!]  
\begin{center}
\epsfig{file=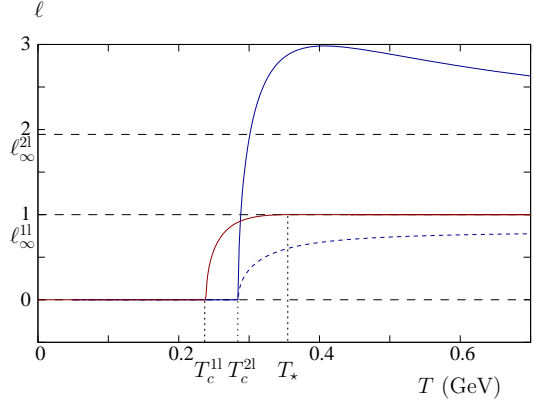,width=7cm}
 \caption{Temperature dependence of the Polyakov loop at leading (red) and next-to-leading (blue) orders in the SU($2$) theory. The horizontal dashed lines denote the corresponding asymptotic values at high temperature, denoted here by $\ell_\infty^{1l}$ and $\ell_\infty^{2l}$, respectively. The respective one- and two-loop critical temperatures are indicated by vertical dashed lines. We have also indicated the temperature $T_\star$ at which the leading-order Polyakov loop reaches its asymptotic value.}\label{fig:pl}
\end{center}
\end{figure}

Such a calculation is of interest, first, in assessing the convergence of the present perturbative approach and, second, because the two-loop contributions resolve some unphysical artifacts of the leading-order results. The first aspect is illustrated in \Fig{fig:pl}, which shows the temperature dependence of the Polyakov loop at one- and two-loop orders. For illustration, the next-to-leading-order expression reads
\beq
\ell(T)=\cos\left(\frac{r_{\rm min}(T)}{2}\right)\left\{1+\frac{g^2 m}{T}\left[\frac{3}{32\pi}+\frac{a(T,r_{\rm min}(T))}{4\pi^2}\sin^2\left(\frac{r_{\rm min}(T)}{2}\right)\right]\right\}
\eeq
where $r_{\rm min}(T)$ is the absolute minimum of the two-loop background field potential and
\beq
a(T,r)=\int_0^\infty\frac{k^2dk}{m^3}\left\{\frac{1}{\cosh(\beta k)-\cos r}-\frac{k^2/\varepsilon^2_k}{\cosh(\beta\varepsilon_k)-\cos r}\right\}\ge0.
\eeq
The critical temperature now depends on both the effective gluon mass and the coupling. We estimate the latter using the same strategy as before: we fit the lattice data for the vacuum ghost and gluon propagators in the Landau gauge against the perturbative expressions at vanishing background field at the appropriate order of approximation.\footnote{We employ the same renormalization conditions as in Ref.~\cite{Tissier_10}, where such fits have been performed. The best fits are obtained with $m=680$~MeV and $g=7.5$ at a scale $\mu=1$~GeV. As a rough error estimation, we have checked that a $30\%$ change in the value of the coupling results in a $10\%$ change in the critical temperature.} We obtain a $20\%$ correction for the critical temperature: $T_c\approx 284$~MeV. This is summarized in the following Table.

\begin{table}[h!]
  \centering
  \begin{tabular}{|l|c|c|c|c|c|}
\hline
$T_c$ (MeV)&one loop \cite{Reinosa:2014ooa}&two loop \cite{Reinosa:2014zta}&lattice \cite{Lucini:2012gg}&FRG/DSE \cite{Fister:2013bh}\\
\hline
SU($2$)&238&284&295&230 (300)\\
\hline
SU($3$)&185&work in progress&270&275\\
\hline
  \end{tabular}
\end{table}
 
\begin{figure}[t!]  
\begin{center}
\epsfig{file=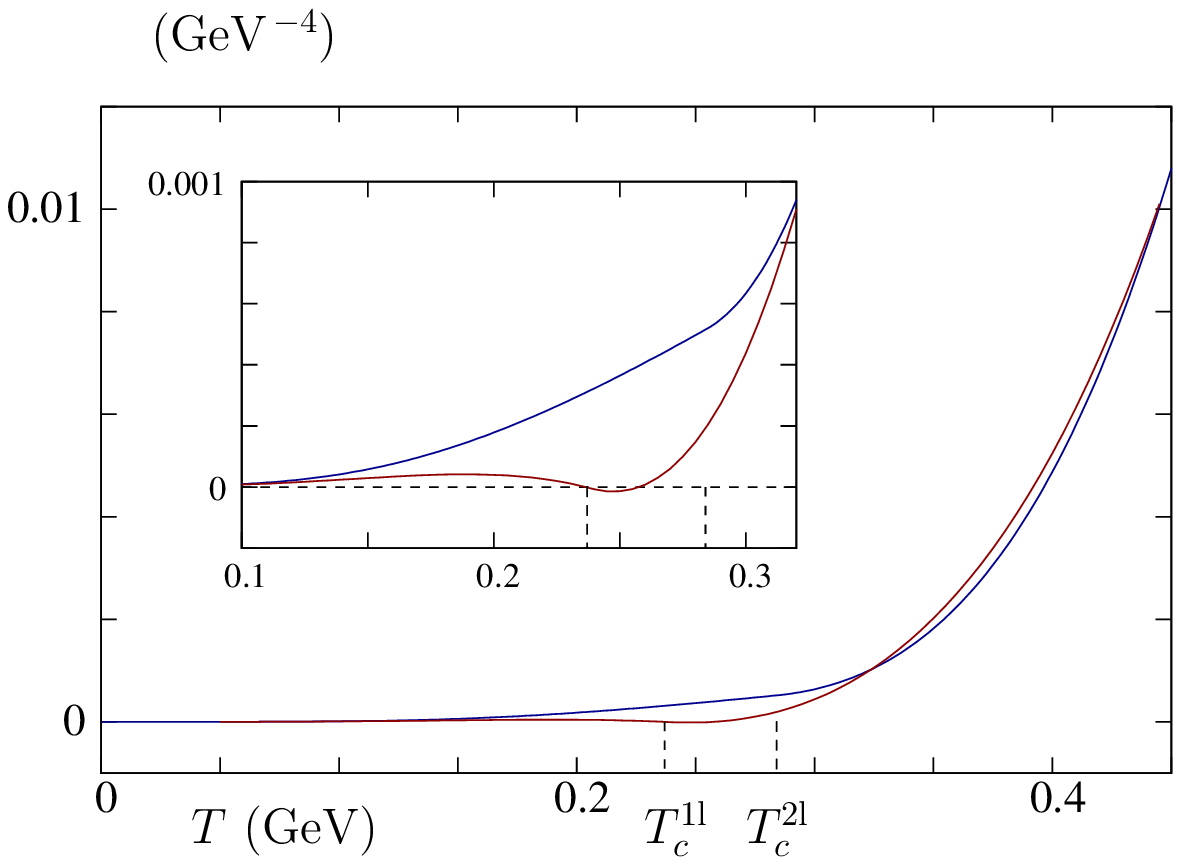,width=6.5cm}\qquad\quad
 \epsfig{file=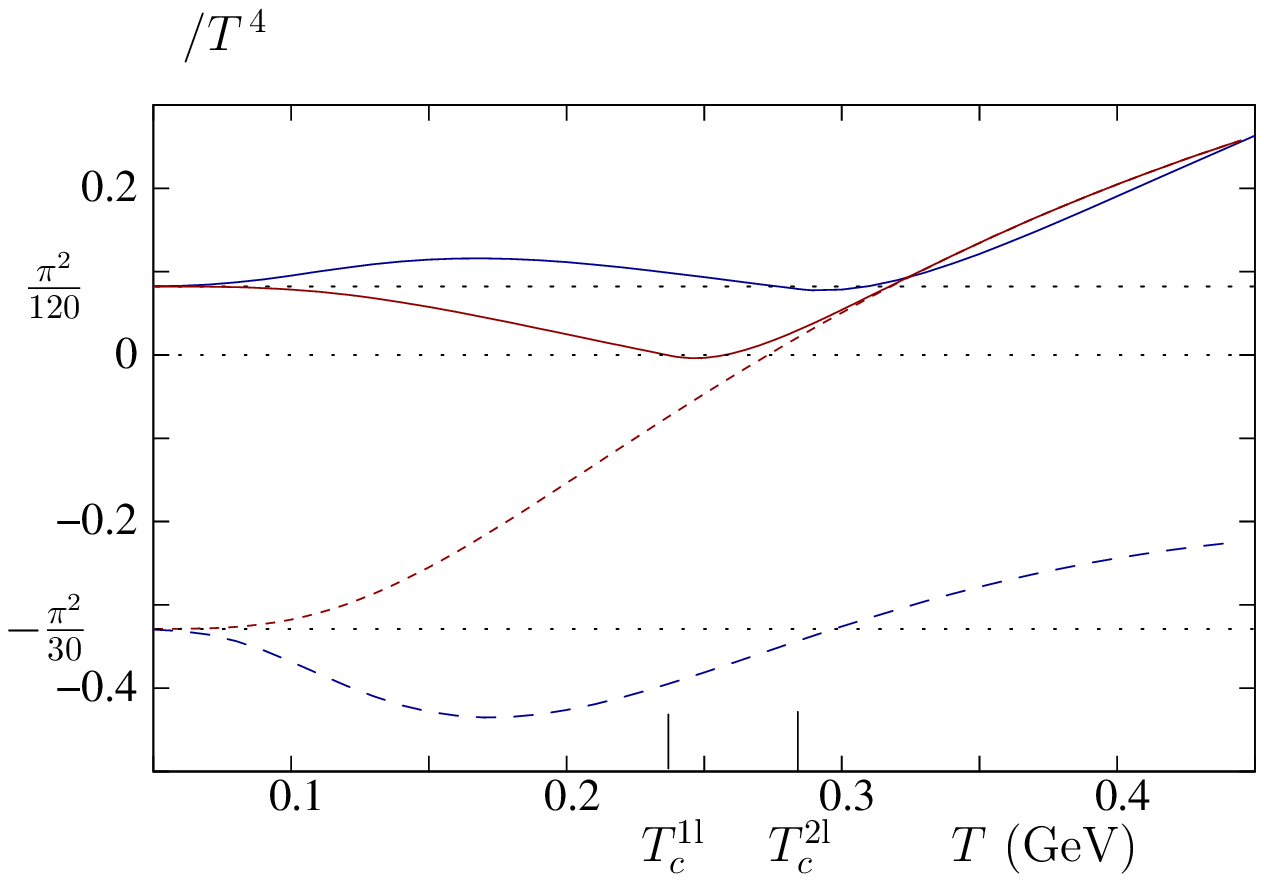,width=6.5cm}
\caption{Left: Thermodynamic pressure at one-loop (red) and two-loop (blue) orders, obtained from the minimum of the background field potential as a function of the temperature. The plot in the inset is a zoom on the low temperature region. The respective one- and two-loop critical temperatures are indicated by vertical dashed lines. Right: Comparison between one-loop (red) and two-loop (blue) results in the massive Landau-DeWitt gauge (plain) and in the massive Landau gauge (dashed).}\label{fig:pressure}
\end{center}
\end{figure}

As for the second aspect mentioned above, there are indeed various unphysical artifacts of the one-loop results which disappear at two loop. The first one is the fact that the leading-order Polyakov loop reaches its asymptotic high temperature value $\ell_\infty^{1l}=1$ at a finite temperature $T_\star$, see \Fig{fig:pl}, which results in a spurious singularity in thermodynamic quantities, as discussed in Ref.~\cite{Reinosa:2014zta}. Such a singularity is not present at two-loop order, where the Polyakov loop only reaches its high temperature value\footnote{The high temperature limit considered here is only formal in the sense that it ignores important physical effects such as the running of the coupling or the physics of hard thermal loops.} $\ell_\infty^{2l}$ asymptotically. Another example concerns the calculation of thermodynamic quantities such as the pressure $p(T)=-V(T,r_{\rm min}(T))$ or the entropy density $s(T)=dp/dT$; see \Fig{fig:pressure}. At leading order, we find a narrow range of temperatures around $T_c^{1l}$ where $p$ and/or $s$ are negative. Again, both unphysical behaviors are cured by the two-loop contributions which are important near the critical temperature, as one might expect. This is discussed in detail in Ref.~\cite{Reinosa:2014zta}. 

It is also interesting to compare the present calculation of the pressure with a nontrivial background field to that at vanishing background, which is equivalent to the (massive) Landau gauge. Figure~\ref{fig:pressure} shows that in the latter case, the leading-order pressure is always negative below a certain temperature\footnote{A similar observation has been made in a different context in Ref.~\cite{Sasaki:2012bi}.} and that the two-loop corrections make the problem even worth since the pressure seems always negative. Clearly, neglecting the nontrivial background in the low temperature regime amounts to expand around an unstable point. This suggests that the strong systematic effects observed in lattice calculations of the Landau gauge gluon propagator near criticality \cite{Mendes:2014gva} might be resolved by performing lattice calculations in the Landau-DeWitt gauge, e.g., along the lines of Ref. \cite{Cucchieri:2012ii}.

Finally, we mention that there remain unphysical contributions to thermodynamic quantities from the massless (e.g., ghost) modes of the gauge-fixed theory. For instance, these leads to spurious $T^4$ contributions to the pressure at low temperatures, as illustrated in \Fig{fig:pressure}. If, as we have discussed above, these modes play a crucial role in ensuring a confining potential for the Polyakov loop at low temperatures, they should not directly contribute to physical quantities such as the pressure or the entropy. At high temperatures, the contributions from unphysical modes cancel out thanks to the effective restoration of the BRST symmetry. However, the cancelation is only incomplete at low temperatures, where the effective gluon mass term cannot be neglected. This important open conceptual issue requires further investigation. An interesting proposal in this direction has been made in Ref.~\cite{Schaden:2014bea} in the context of the Gribov-Zwanziger approach.

\section{Conclusions and perspectives}

In conclusion, we have proposed a modified perturbative approach to the physics of the static quark confinement-deconfinement transition that takes into account the Gribov issue in (gauge-fixed) continuum calculations in an effective way. This is based on a massive extension of the Landau-DeWitt gauge in the context of background field methods. The approach correctly captures the physics of the phase transition in Yang-Mills theories already at leading order in perturbation theory and next-to-leading-order corrections clearly  improve the results.  Although there are still important open questions, in particular, concerning thermodynamic quantities at low temperatures, we believe this is an interesting step towards a (semi)analytical understanding of the dynamics at work in the transition region. 

There are obvious extensions of the work presented here, among which, the calculation of the two-loop contributions in the SU($3$) theory, the application of our approach at finite (real or imaginary) chemical potential, or the calculation of the finite temperature ghost and gluon propagators at one-loop order in presence of the nontrivial background field. These are works in progress.


\begin{thebibliography}{99}

\bibitem{Petreczky:2012rq}
  P.~Petreczky,
  J.\ Phys.\ G {\bf 39} (2012) 093002;
  O.~Philipsen,
  Prog.\ Part.\ Nucl.\ Phys.\  {\bf 70} (2013) 55.

\bibitem{Borsanyi-thisproc}
 Sz. Bors\'anyi, these proceedings.
 
\bibitem{Fischer:2009wc}
  C.~S.~Fischer,
  Phys.\ Rev.\ Lett.\  {\bf 103} (2009) 052003.
 
\bibitem{Fister:2013bh} 
  J.~Braun, H.~Gies and J.~M.~Pawlowski,
  Phys.\ Lett.\ B {\bf 684} (2010) 262;
  L.~Fister and J.~M.~Pawlowski,
  Phys.\ Rev.\ D {\bf 88}, 045010 (2013).
 
\bibitem{Fisher-thisproc}
 See the contributions of C. Fisher, J. Pawlowski, and Y. Liu to these proceedings and references therein.
 
 
\bibitem{Pisarski:2000eq}
  R.~D.~Pisarski,
  Phys.\ Rev.\ D {\bf 62} (2000) 111501.

\bibitem{Dumitru:2012fw} 
  A.~Dumitru, Y.~Guo, Y.~Hidaka, C.~P.~K.~Altes and R.~D.~Pisarski,
  Phys.\ Rev.\ D {\bf 86}, 105017 (2012).
  
\bibitem{Gribov77}
  V.~N.~Gribov,
  Nucl.\ Phys.\  B {\bf 139} (1978) 1;
  I.~M.~Singer,
  Commun.\ Math.\ Phys.\  {\bf 60} (1978) 7.

\bibitem{Boucaud:2011ug}
  P.~Boucaud, J.~P.~Leroy, A.~L.~Yaouanc, J.~Micheli, O.~Pene and J.~Rodriguez-Quintero,
  Few-Body Syst.\  {\bf 53} (2012) 387;
  A.~Maas,
  Phys.\ Rep.\  {\bf 524} (2013) 203.

\bibitem{Neuberger:1986vv}
  H.~Neuberger,
  Phys.\ Lett.\  B {\bf 175}, 69 (1986);
  Phys.\ Lett.\  B {\bf 183} (1987) 337.
  
\bibitem{Zwanziger89}
  D.~Zwanziger,
  Nucl.\ Phys.\  B {\bf 323}, 513 (1989);
  Nucl.\ Phys.\  B {\bf 399}, 477 (1993).

\bibitem{Burgio:2008jr}
  G.~Burgio, M.~Quandt and H.~Reinhardt,
  Phys.\ Rev.\ Lett.\  {\bf 102} (2009) 032002.

\bibitem{Dudal08}
  D.~Dudal, J.~A.~Gracey, S.~P.~Sorella, N.~Vandersickel and H.~Verschelde,
  Phys.\ Rev.\  D {\bf 78} (2008) 065047.
  
  
\bibitem{Zwanziger:2004np}
  D.~Zwanziger,
  Phys.\ Rev.\ Lett.\  {\bf 94} (2005) 182301;
  Phys.\ Rev.\ D {\bf 76} (2007) 125014.

\bibitem{Fukushima:2013xsa}
  K.~Fukushima and N.~Su,
  Phys.\ Rev.\ D {\bf 88} (2013) 076008.

\bibitem{Canfora:2013kma}
  F.~Canfora, P.~Pais and P.~Salgado-Rebolledo,
  Eur.\ Phys.\ J.\ C {\bf 74} (2014) 2855.

  
  
  
   \bibitem{Tissier_10}
   M.~Tissier and N.~Wschebor,
   Phys.\ Rev.\  D {\bf 82} (2010) 101701;   Phys.\ Rev.\  D {\bf 84} (2011) 045018.

\bibitem{Weber:2012vf}
  A.~Weber,
  J.\ Phys.\ Conf.\ Ser.\  {\bf 378} (2012) 012042.

  \bibitem{Curci76}
  G.~Curci and R.~Ferrari,
  Nuovo Cim.\ A {\bf 32}, 151 (1976).
 
\bibitem{Pelaez:2013cpa}
  M.~Pel\'aez, M.~Tissier and N.~Wschebor,
  Phys.\ Rev.\ D {\bf 88} (2013) 125003.

\bibitem{Pelaez:2014mxa}
  M.~Pel\'aez, M.~Tissier and N.~Wschebor,
  Phys.\ Rev.\ D {\bf 90} (2014) 065031.

\bibitem{Reinosa:2013twa} 
  U.~Reinosa, J.~Serreau, M.~Tissier and N.~Wschebor,
  Phys.\ Rev.\ D {\bf 89}, 105016 (2014).

\bibitem{Serreau:2012cg}
  J.~Serreau and M.~Tissier,
  Phys.\ Lett.\ B {\bf 712} (2012) 97.
  
\bibitem{Serreau:2013ila}
  J.~Serreau, M.~Tissier and A.~Tresmontant,
  Phys.\ Rev.\ D {\bf 89} (2014) 125019.
  
\bibitem{Kosterlitz:1973xp}
 J.~M.~Kosterlitz and D.~J.~Thouless,
 J.\ Phys.\ C {\bf 6} (1973) 1181;
 J.~V.~Jose, L.~P.~Kadanoff, S.~Kirkpatrick and D.~R.~Nelson,
 Phys.\ Rev.\ B {\bf 16}, 1217 (1977).
  
\bibitem{Reinosa:2014ooa}
  U.~Reinosa, J.~Serreau, M.~Tissier and N.~Wschebor,
  Phys.\ Lett.\ B {\bf 742} (2015) 61.
  
\bibitem{Reinosa:2014zta}
  U.~Reinosa, J.~Serreau, M.~Tissier and N.~Wschebor,
  Phys.\ Rev.\ D {\bf 91} (2015) 4,  045035.

\bibitem{Weinberg:1996kr} 
  S.~Weinberg,
 ``The quantum theory of fields. Vol. 2: Modern applications,''
  Cambridge, UK: Univ. Pr. (1996).
  
\bibitem{Weiss:1980rj}
  N.~Weiss,
  Phys.\ Rev.\ D {\bf 24} (1981) 475;
  D.~J.~Gross, R.~D.~Pisarski and L.~G.~Yaffe,
  Rev.\ Mod.\ Phys.\  {\bf 53} (1981) 43.

\bibitem{Lucini:2012gg}
  B.~Lucini and M.~Panero,
  Phys.\ Rept.\  {\bf 526} (2013) 93.

\bibitem{Gupta:2007ax}
  S.~Gupta, K.~Huebner and O.~Kaczmarek,
  Phys.\ Rev.\ D {\bf 77} (2008) 034503.

\bibitem{Mykkanen:2012ri}
  A.~Mykkanen, M.~Panero and K.~Rummukainen,
  JHEP {\bf 1205} (2012) 069.

\bibitem{Sasaki:2012bi}
  C.~Sasaki and K.~Redlich,
  Phys.\ Rev.\ D {\bf 86} (2012) 014007.
  
\bibitem{Mendes:2014gva}
  T.~Mendes and A.~Cucchieri,
  PoS LATTICE {\bf 2013} (2014) 456.

\bibitem{Cucchieri:2012ii}
  A.~Cucchieri and T.~Mendes,
  Phys.\ Rev.\ D {\bf 86} (2012) 071503.

\bibitem{Schaden:2014bea}
  M.~Schaden and D.~Zwanziger,
  arXiv:1412.4823 [hep-ph].
  
  
  

\end{thebibliography}
\end{document}